\documentclass[11pt,a4paper]{article}

\usepackage{graphics}{}
\pdfoutput=1
\usepackage[pdftex]{}
\usepackage{amsmath,amssymb,amsfonts}

\usepackage[dvips]{epsfig}
\usepackage{multicol}
\usepackage{xspace}
\usepackage{url}
\usepackage{hyperref}
\usepackage{amsmath}

\begin{document}

\vspace{1.8cm}
\begin{center}
{\Large \bf NA49/NA61: results and plans on beam energy and system size scan
at the CERN SPS\footnote{
invited talk at the Quark Matter 2011, Annecy, France}
}
\end{center}

\vspace*{0.8cm}
\begin{center}
Marek Gazdzicki$^{a,b}$ for the NA49 and NA61/SHINE Collaborations 
\end{center}

\vspace*{0.2cm}
\begin{center}
{\it $^{a}$Goethe--University  Frankfurt, Germany\\}
{\it $^{b}$Jan Kochanowski University, Kielce, Poland\\}
\end{center}

\vspace*{1.0cm}
\begin{abstract}
{\small
This paper presents results and plans of the NA49 and
NA61/SHINE experiments at the CERN Super Proton Synchrotron
concerning the study of relativistic nucleus-nucleus  interactions.
First, the NA49 evidence for the energy threshold of creating
quark-gluon plasma, the onset of deconfinement,
in central lead-lead collisions around 30$A$~GeV is reviewed.
Then the status of the NA61/SHINE systematic  study of properties
of the onset of deconfinement is presented.
Second, the search for the critical point of strongly interacting matter
undertaken by both experiments is discussed.
NA49 measured large fluctuations at the top SPS energy, 158$A$~GeV,
in collisions of light and medium size nuclei. They seem to indicate
that the critical point exists and is located close to baryonic chemical
potential of about 250~MeV. 
The NA61/SHINE beam energy and system size scan started in 2009
will provide evidence for the existence of the critical point or
refute the interpretation of the NA49 fluctuation data in terms of
the critical point.
}
\vspace*{2.0cm}
\end{abstract}

\thispagestyle{empty}

\newpage
\begin{center}
The NA61/SHINE Collaboration
\end{center}

\noindent
N.~Abgrall${}^{22}$,
A.~Aduszkiewicz${}^{23}$,
B.~Andrieu${}^{11}$,
T.~Anticic${}^{13}$,
N.~Antoniou${}^{18}$,
J.~Argyriades${}^{22}$,
A.~G.~Asryan${}^{15}$,
B.~Baatar${}^{9}$,
A.~Blondel${}^{22}$,
J.~Blumer${}^{5}$
% ...

\vspace{0.5cm}
\noindent
% ...

\section{Introduction}

In 1999
the NA49 experiment at the CERN Super Proton Synchrotron
started a search for the onset of quark-gluon plasma (QGP)~\cite{qgp}
creation with
data taking for central Pb+Pb collisions at 40$A$~GeV.
Runs at 80$A$ and 20$A$,  30$A$~GeV followed in 2000 and 2002, respectively.
This search was motivated by predictions of a statistical model
of the early stage of nucleus--nucleus collisions~\cite{GaGo}
that the onset of deconfinement should lead to rapid 
changes of the energy dependence numerous hadron production
properties, all appearing in a common energy domain.
The conjectured features were observed~\cite{evidence,review}
around 30$A$~GeV
and dedicated experiments, NA61/SHINE at the CERN SPS and the
Beam Energy Scan at BNL RHIC, continue detailed studies in
the energy region of the onset of deconfinement. 
 
Hadrons produced in collisions of light and medium size nuclei
at collision energies higher than the energy of the onset
of deconfinement 
may freeze-out just below the transition
line between hadron gas and QGP. Thus  their production
properties may be sensitive to properties of the transition.
In particular, freeze-out in the vicinity of 
the  critical point may lead to a characteristic
pattern of event-by-event fluctuations~\cite{cp}.
Motivated by these predictions NA49 and NA61/SHINE at the CERN SPS
as well as STAR and PHENIX at the BNL RHIC have started
a systematic search for the critical point.
\begin{figure}[!h]
\begin{center}
\begin{minipage}[b]{0.8\linewidth}
\includegraphics[width=1.0\linewidth]{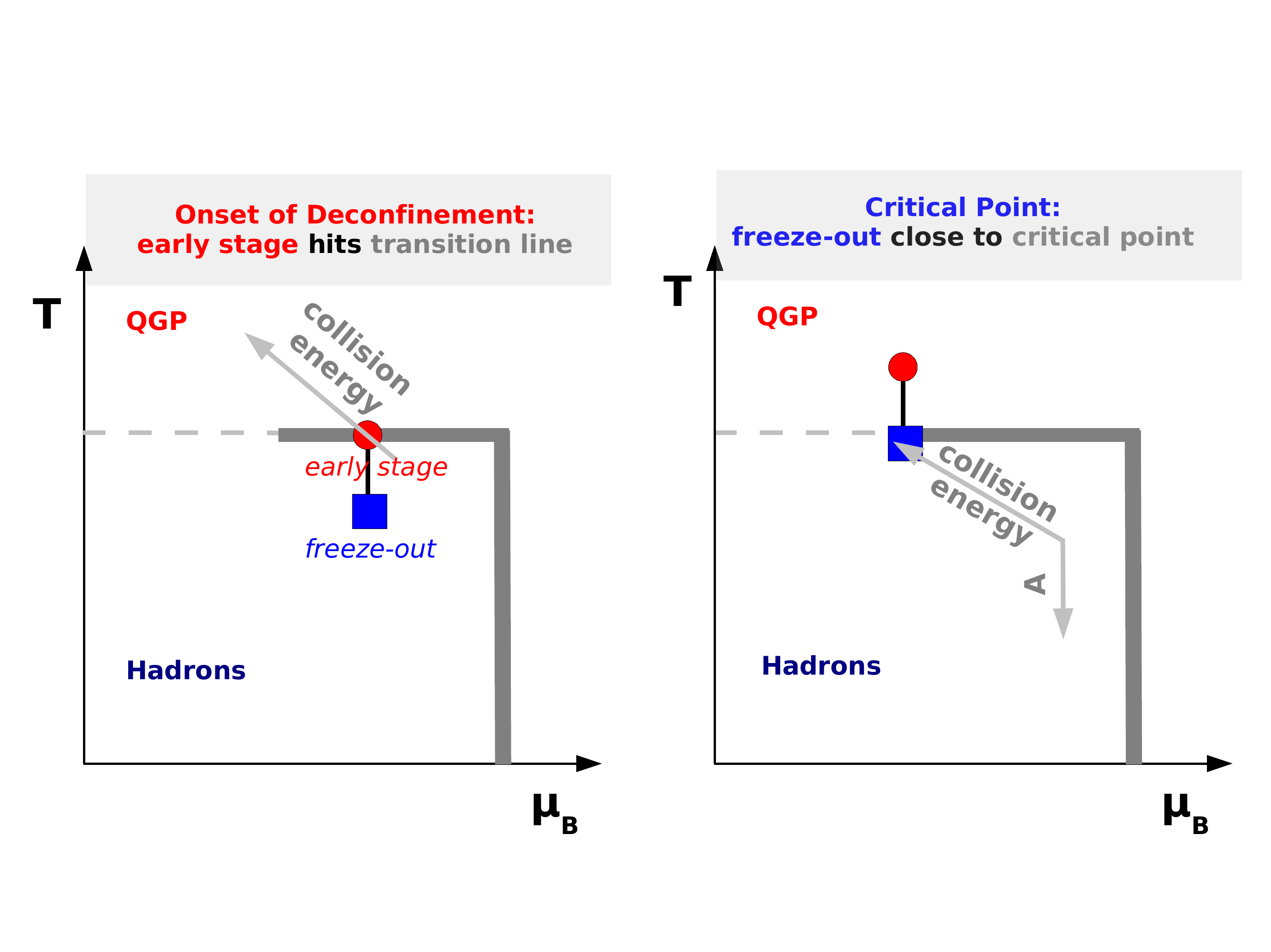}
\end{minipage}
\vspace*{-0.5cm}
\caption{\label{cpod}
Sketches illustrating the basic ideas behind the experimental
search for and the study of the
onset of deconfinement ({\it left}) and the critical point ({\it right}).
The transition line between hadron gas and quark-gluon plasma
is indicated by solid (1$^{st}$ order transition) and
dotted (cross-over) lines, which meet at the critical point
(2$^{nd}$ order transition). 
The parameters of matter created in A+A collisions at the  
early stage and at freeze-out are indicated by the closed circle
and square, respectively. Their dependence on collision energy and
size of colliding nuclei is shown by arrows.
}
\end{center}
\end{figure}

The two sketches presented in Fig.~\ref{cpod}
illustrate the basic ideas behind the experimental
strategies in the search for and the study of the onset of deconfinement
and the critical point.

In this report results of NA49 and plans of NA61/SHINE
concerning  the onset of deconfinement (Section~2) and
the critical point (Section~3) are briefly presented.

\section{Onset of Deconfinement}

\begin{figure}[!htb]
\begin{center}
\begin{minipage}[b]{0.95\linewidth}
\includegraphics[width=0.5\linewidth]{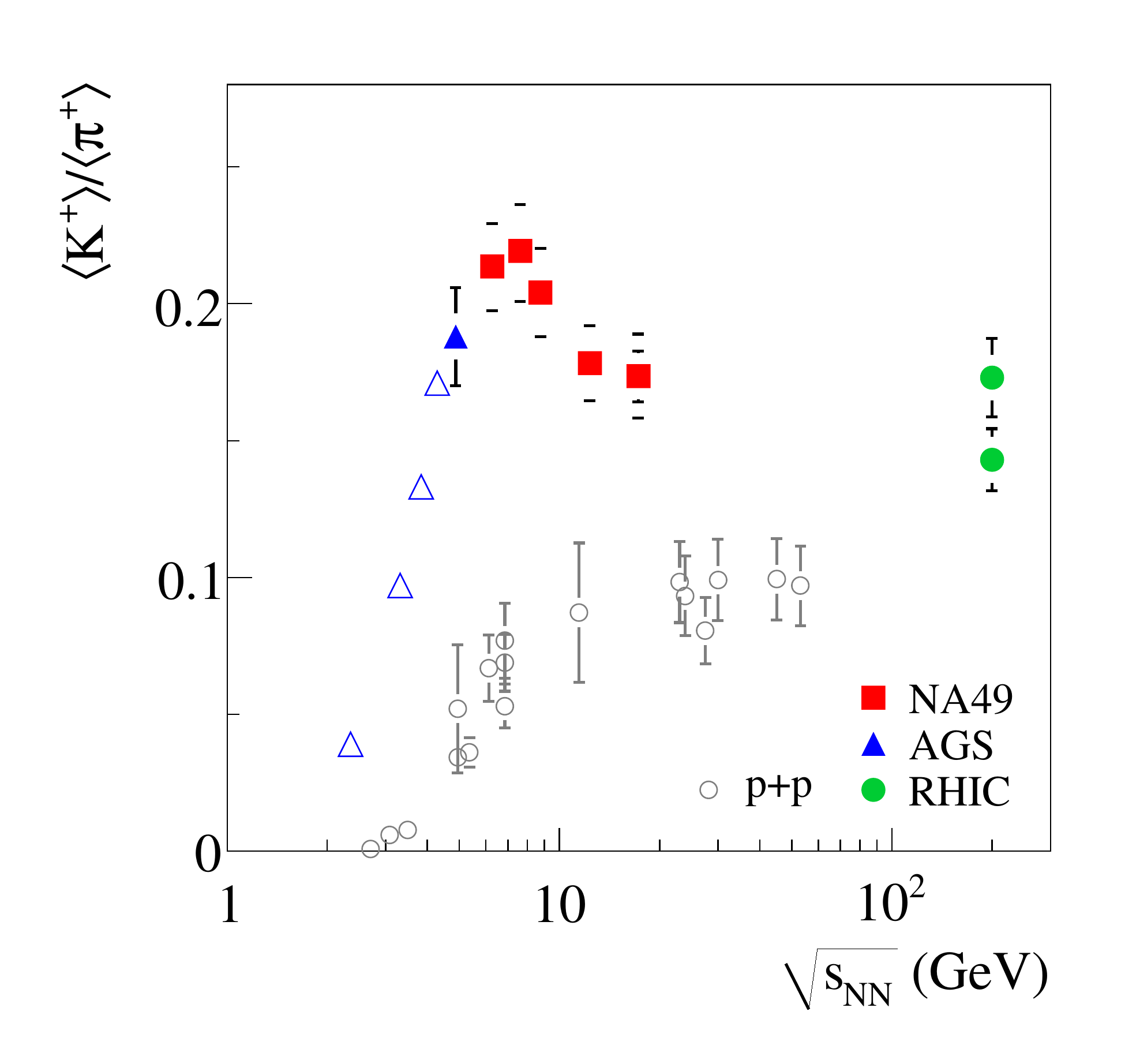}
\includegraphics[width=0.5\linewidth]{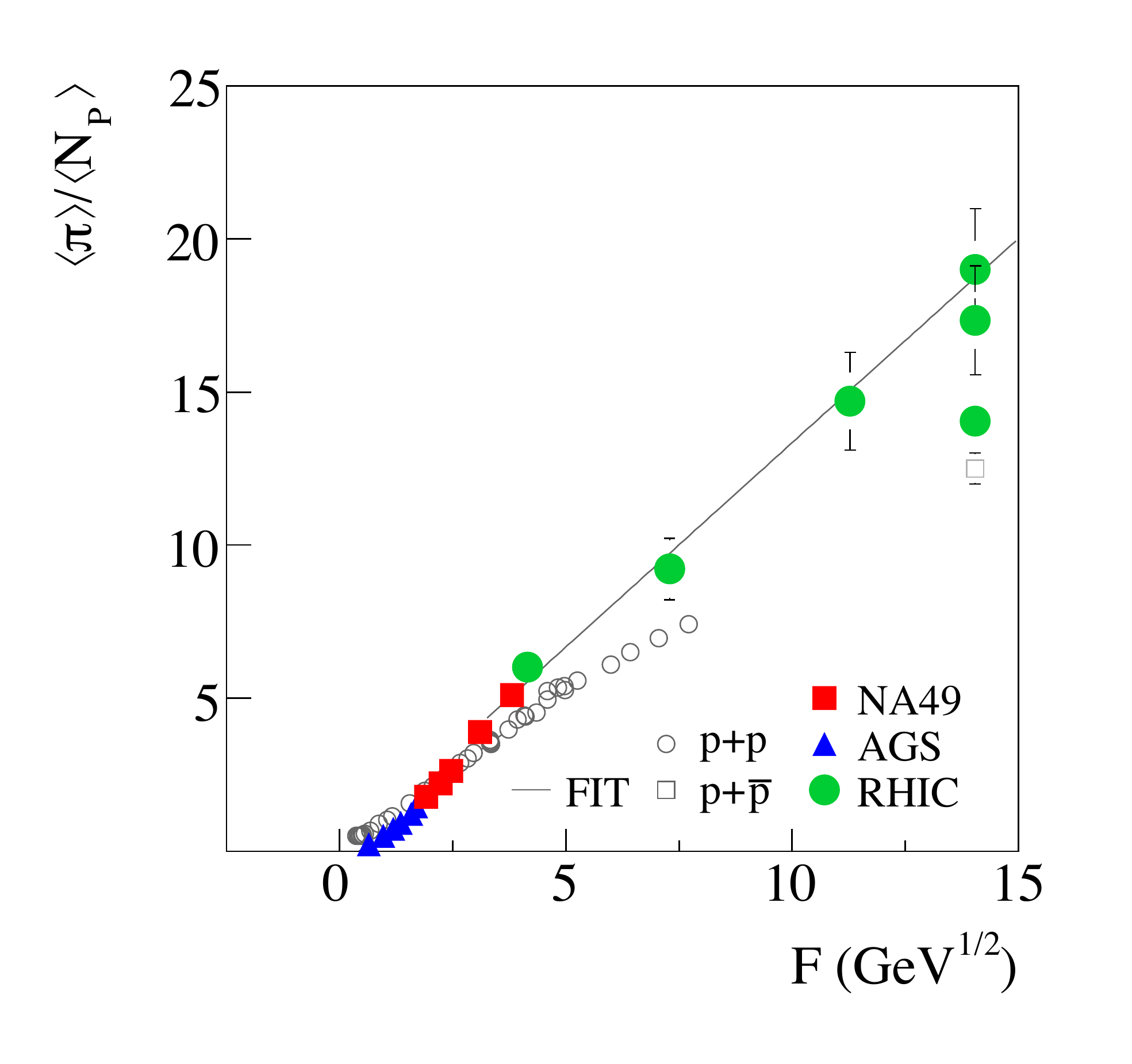}
\includegraphics[width=0.5\linewidth]{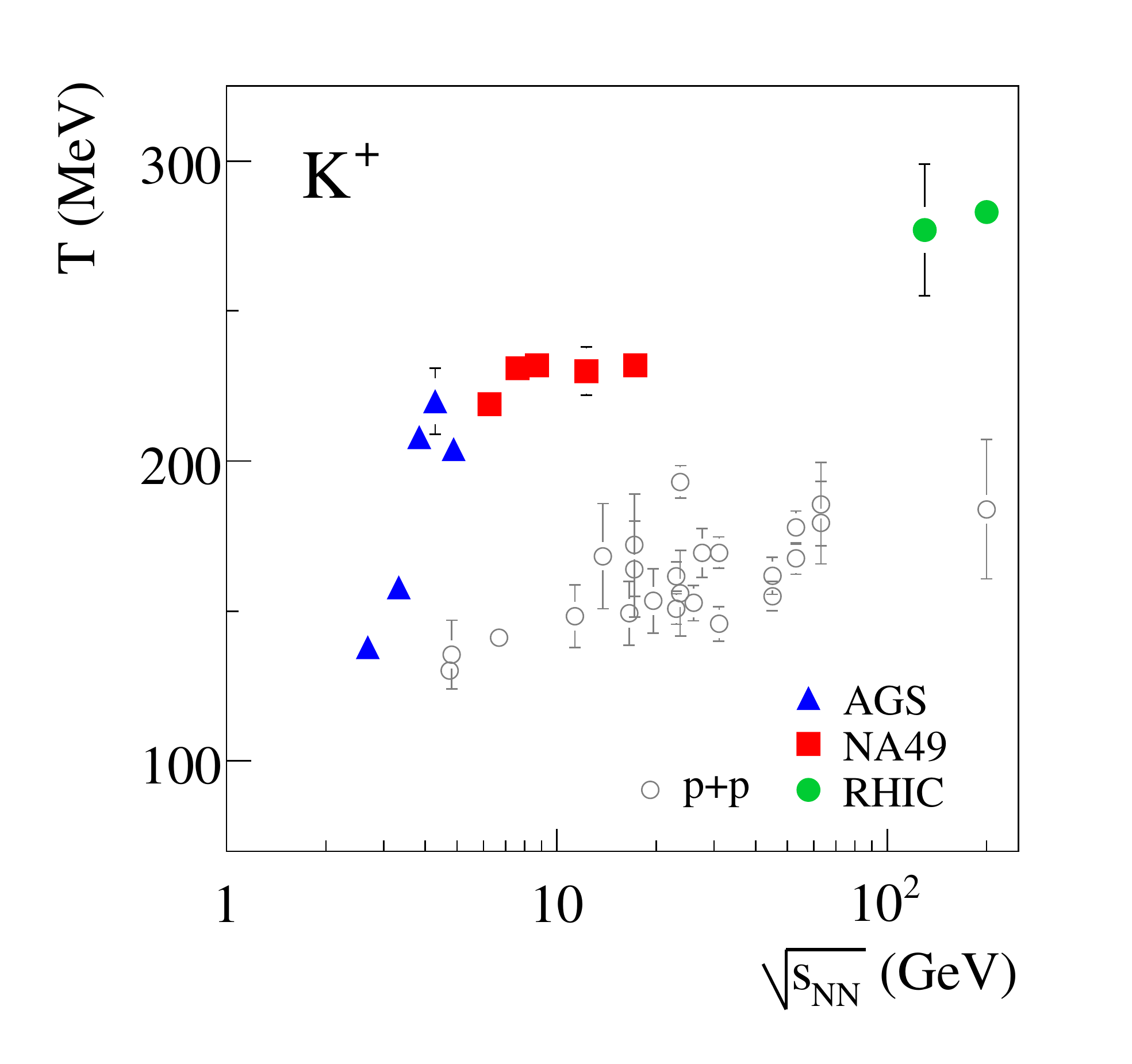}
{\hspace*{0.2 cm} 
\includegraphics[width=0.5\linewidth]{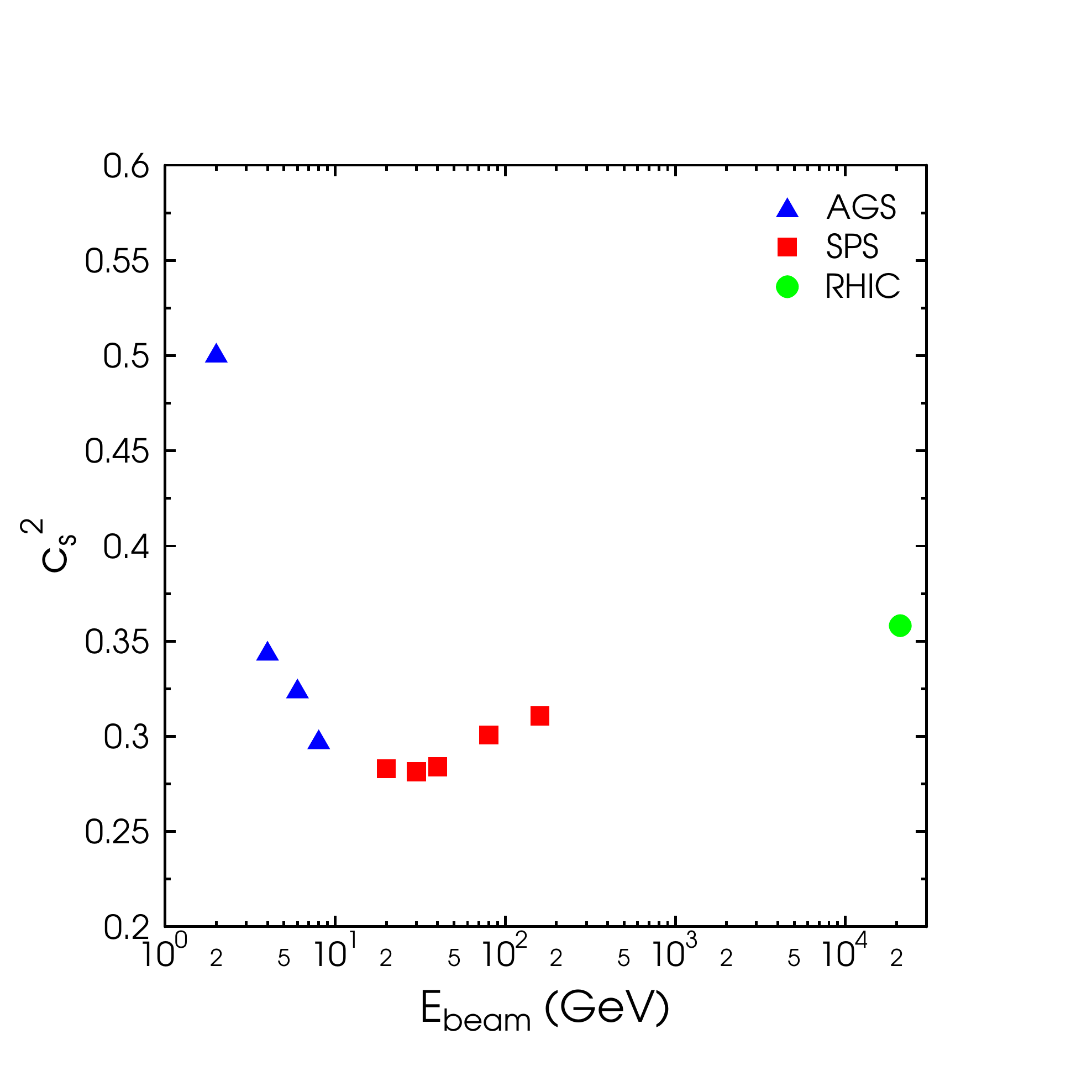}}
\end{minipage}
\caption{\label{heating_curves}
Heating curves of strongly interacting matter.
Hadron production properties (see Ref.~\cite{review} for details) 
are plotted
as a function of collision energy for central Pb+Pb (Au+Au)
collisions and p+p interactions
(open circles).
The observed changes of the energy dependence for central Pb+Pb (Au+Au)
collisions are related to: 
decrease of the mass of strangeness carries and the ratio of
strange to non-strange $dof$ ({\it horn}: top-left plot),
increase of entropy production ({\it kink}: top-right plot),
weakening of transverse ({\it step}: bottom-left plot)  
and longitudinal ({\it dale}: bottom-right plot) 
expansion at the onset of deconfinement.  
}
\end{center}
\end{figure}

The detailed review of the experimental and theoretical
status of the NA49 evidence for the onset of deconfinement
can be found in the recent review~\cite{review}.
The evidence is based on the  observation
that numerous hadron production properties measured
in central Pb+Pb collisions change their energy dependence
in a common energy domain (starting from about 30$A$~GeV) and
that these changes are consistent with the predictions
for the onset of deconfinement. The four representative
plots  with the structures
referred to as $horn$, $kink$, $step$ and $dale$~\cite{review}
are shown in Fig.~\ref{heating_curves}.

\begin{figure}[!htb]
\begin{center}
\begin{minipage}[b]{0.9\linewidth}
\includegraphics[width=0.5\linewidth]{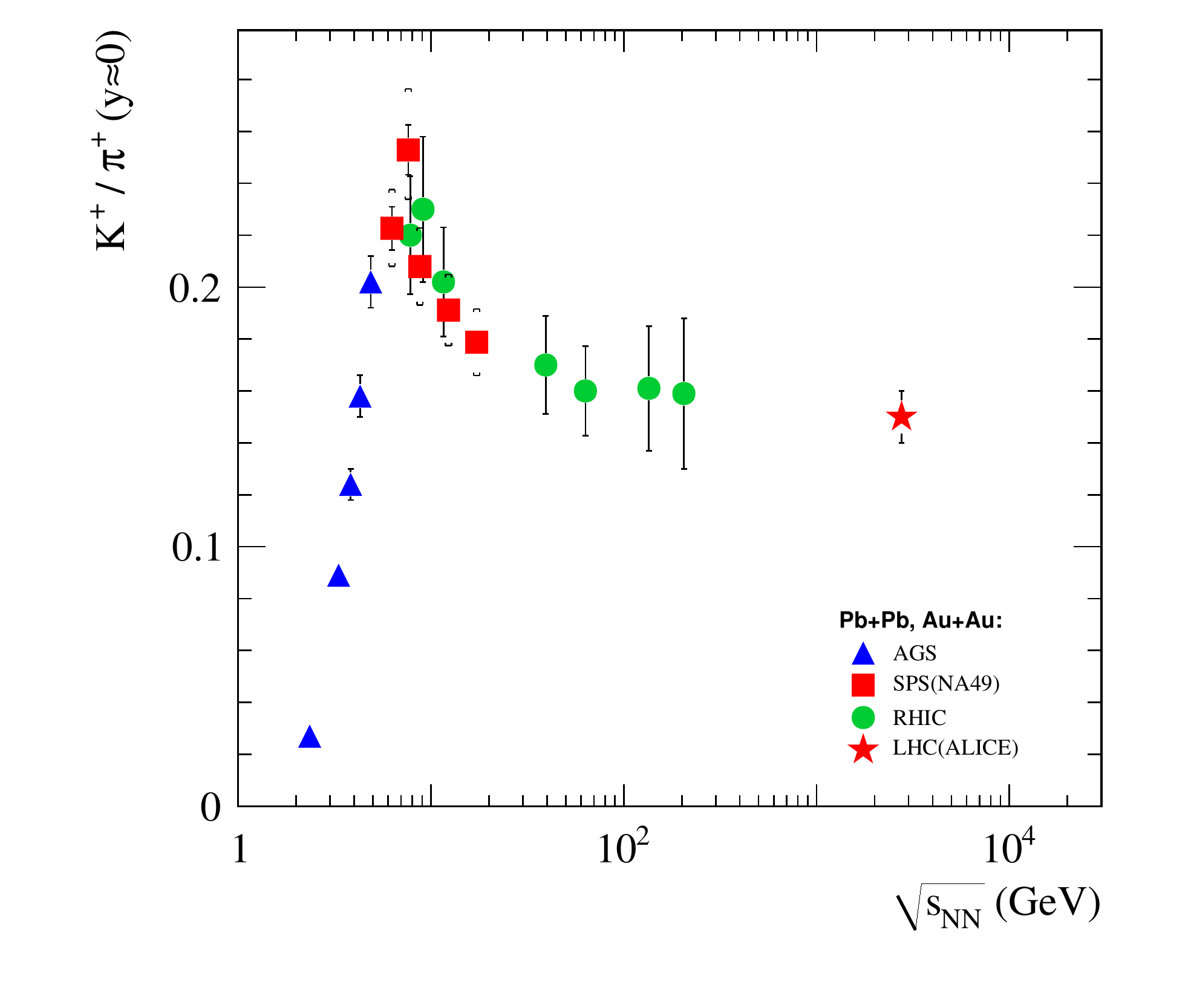}
\includegraphics[width=0.5\linewidth]{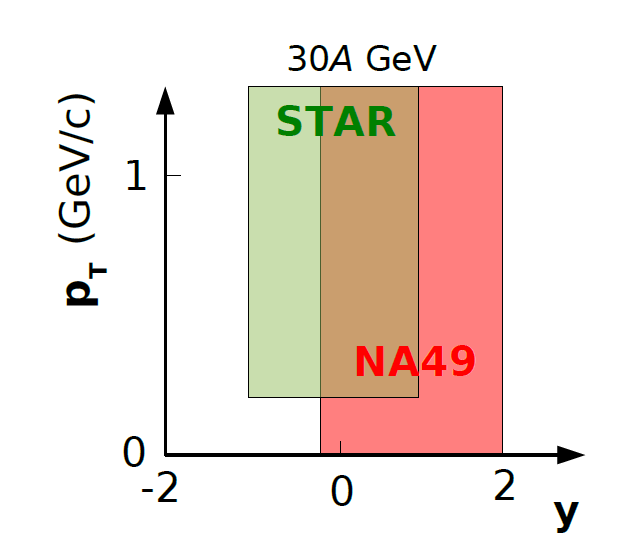}
\end{minipage}
\caption{\label{new_horn}
$Left:$ Energy dependence of the K$^+$/$\pi^+$ ratio at mid-rapidity
for central Pb+Pb (Au+Au) collisions
showing also the new RHIC and LHC data.
$Right:$ Schematic comparison of the NA49 and STAR acceptance for measurements
of identified hadrons at 30$A$~GeV in transverse momentum (p$_T$) and rapidity (y).
}
\end{center}
\end{figure}

At Quark Matter 2011 new data from the RHIC beam energy scan 
with gold nuclei~\cite{kumar}
and the Large Hadron Collider 2010 run 
with lead nuclei at 2.76~TeV~\cite{schukraft}
were presented. They strongly support the NA49  evidence for the onset
of deconfinement.
The RHIC results~\cite{kumar} confirm the NA49 measurements at the onset energies.
The LHC data demonstrate that the energy dependence of hadron production properties
shows rapid changes only at the low SPS energies. A smooth evolution is observed
between the top SPS (17.2~GeV) and the current LHC (2.76~TeV) energies. This agrees with
the interpretation of the NA49 structures as due to the onset of deconfinement
and the expectation of only a smooth evolution of the
quark-gluon plasma properties with increasing collision energy above the onset energy.
As an example the energy dependence of the K$^+$/$\pi^+$ ratio at mid-rapidity
for central Pb+Pb (Au+Au) collisions
with the new RHIC and LHC data is shown in Fig.~\ref{new_horn}~{\it left}.
The STAR and ALICE  measurements
of identified particle spectra  are restricted to the mid-rapidity
region and transverse momenta larger than several hundred~MeV/c.
This is illustrated in Fig.~\ref{new_horn}~{\it right}, where
a schematic comparison of the NA49 and STAR acceptances  at 30$A$~GeV
is shown.
Thus it is important to note that the collider results presented 
in Fig.~\ref{new_horn}~{\it left} include extrapolation to p$_T = 0$
which increases systematic uncertainty of the results.

The observed signals of the onset of deconfinement concern
single particle production properties. The search for signals
in particle correlations and event-by-event fluctuations is
up to now inconclusive~\cite{review}.
Moreover, the
energy dependence of event-by-event K/$\pi$ and K/p fluctuations
measured by NA49 and STAR in central Pb+Pb (Au+Au) collisions,
is different~\cite{tim,terry}. Both collaborations work on clarification of 
the observed differences.               

\begin{figure}[!htb]
\begin{center}
\begin{minipage}[b]{0.7\linewidth}
\includegraphics[width=1.0\linewidth]{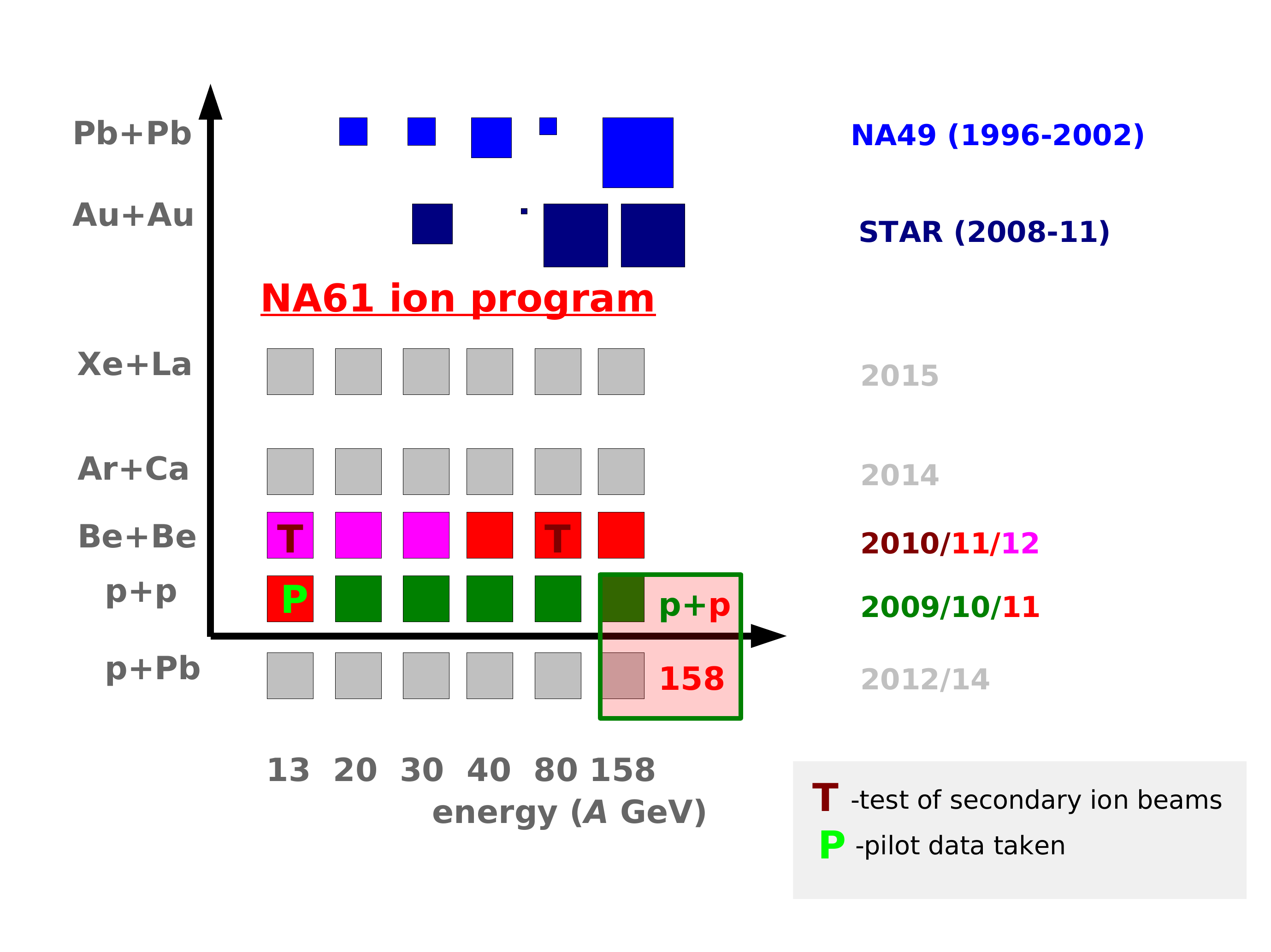}
\end{minipage}
\caption{\label{plans}
The NA61 data taking status and plans. For comparison
data collected by NA49 and STAR are also indicated.
}
\end{center}
\end{figure}

An important part of the ion program of the NA61/SHINE experiment
at the CERN SPS is the study of the properties of the onset of
deconfinement~\cite{proposal}. NA61 aims to establish the system size dependence
of the energy dependence of hadron production properties.
This requires measurements for different pairs of colliding nuclei.
The transition from the structure-less energy dependence for
p+p interactions to the one with the structures related to the
onset of deconfinement is expected to take place for collisions of
medium size nuclei ($A \approx 30-40$). Consequently, in order to localize this 
transition, data on collisions of light nuclei ($A \approx 10$),
medium nuclei ($A \approx 30-40$) and heavy nuclei ($A \ge 100$) 
will be taken in the coming years.
The  program started in 2009 with the energy scan of p+p interactions
which are needed for a precise determination of the p+p baseline.
The NA61 data taking status and plans are presented in Fig.~\ref{plans}.
For comparison
data collected by NA49 and STAR are also indicated.

\begin{figure}[!htb]
\begin{center}
\begin{minipage}[b]{0.8\linewidth}
\includegraphics[width=0.45\linewidth]{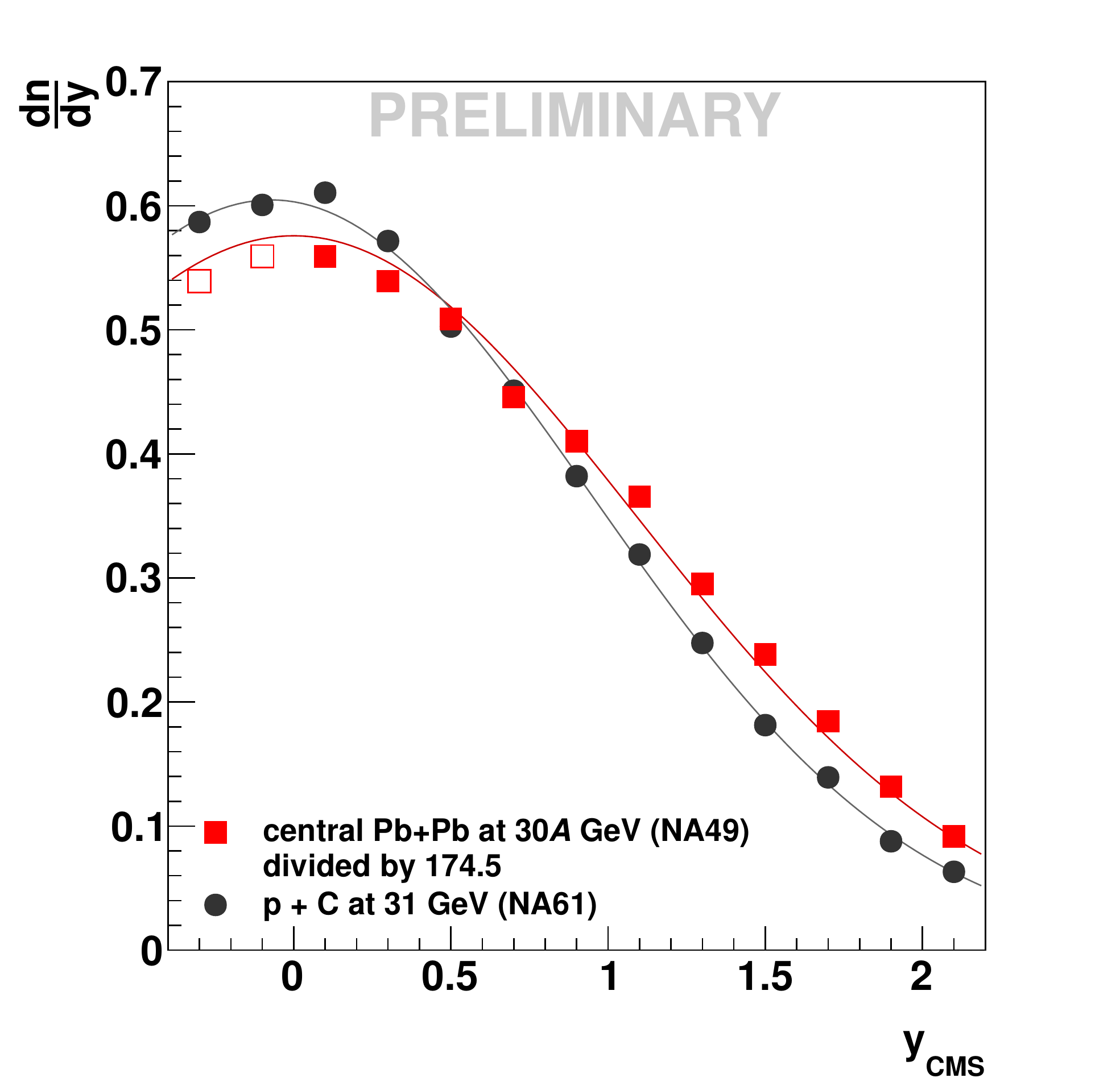}
\includegraphics[width=0.45\linewidth]{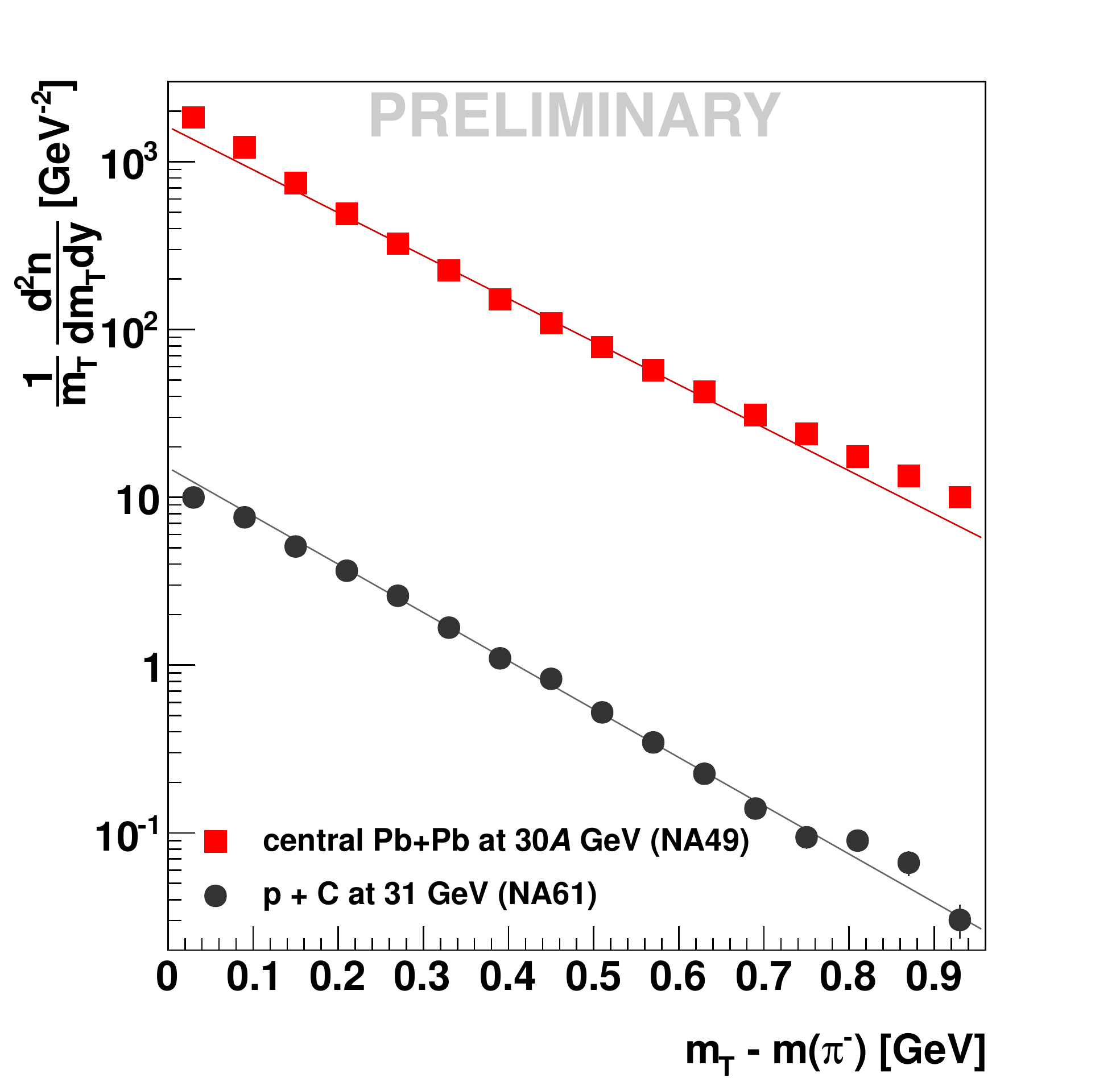}
\end{minipage}
\caption{\label{pions}
Rapidity ({\it left}) and transverse mass spectra at mid-rapidity 
($0 < y < 0.2$) ({\it right})
of negatively charged pions produced in all production p+C interactions
at 31~GeV and central (7\%) Pb+Pb collisions at  30$A$~GeV.
The Pb+Pb rapidity spectrum is divided by the mean number of wounded
nucleons from the projectile nucleus.
}
\end{center}
\end{figure}

First results from NA61 relevant to the study of properties of
the onset of deconfinement are presented in 
Fig.~\ref{pions}~\cite{antoni}.
The rapidity and transverse mass spectra at mid-rapidity of
negatively charged pions in all production p+C interactions
at 31~GeV are compared to the corresponding NA49 results for
central Pb+Pb collisions at 30$A$~GeV~\cite{evidence}.  
The p+C rapidity spectrum is shifted towards target rapidity with
respect to Pb+Pb collisions due to the projectile-target asymmetry of the
initial state. 
The mean pion multiplicity  in the forward hemisphere is approximately 
proportional to the mean number of wounded nucleons of the projectile
nucleus. This reflects previous observations~\cite{review} 
that at the energy of the onset of deconfinement (30$A$~GeV) the mean pion
multiplicity in central Pb+Pb collisions agrees with that predicted
by the Wounded Nucleon Model~\cite{wnm}.  
The shape of the transverse mass spectra at mid-rapidity changes
from a convex form in p+C interactions to
a concave one in central Pb+Pb collisions 
(with respect to the corresponding exponential fits).
Within hydrodynamical approaches
this is due to the significant collective flow in Pb+Pb collisions which is absent
in p+C interactions. 

\section{Critical Point}

The discovery of the onset of deconfinement discussed above 
implies the existence of
QGP and of a transition region between confined and QGP phases.
The most popular possibility concerning the
structure of the transition region~\cite{ssr}, sketched in Fig.~\ref{cpod}, claims
that a 1$^{st}$ order phase transition (thick gray line) separates
both phases in the high baryonic chemical potential domain. In the
low baryonic chemical potential domain a rapid crossover is
expected. The end point of the 1$^{st}$ order phase
transition line is the critical point (the 2$^{nd}$ order phase transition).
\begin{figure}[!htb]
\begin{center}
\begin{minipage}[b]{0.7\linewidth}
\includegraphics[width=1.0\linewidth]{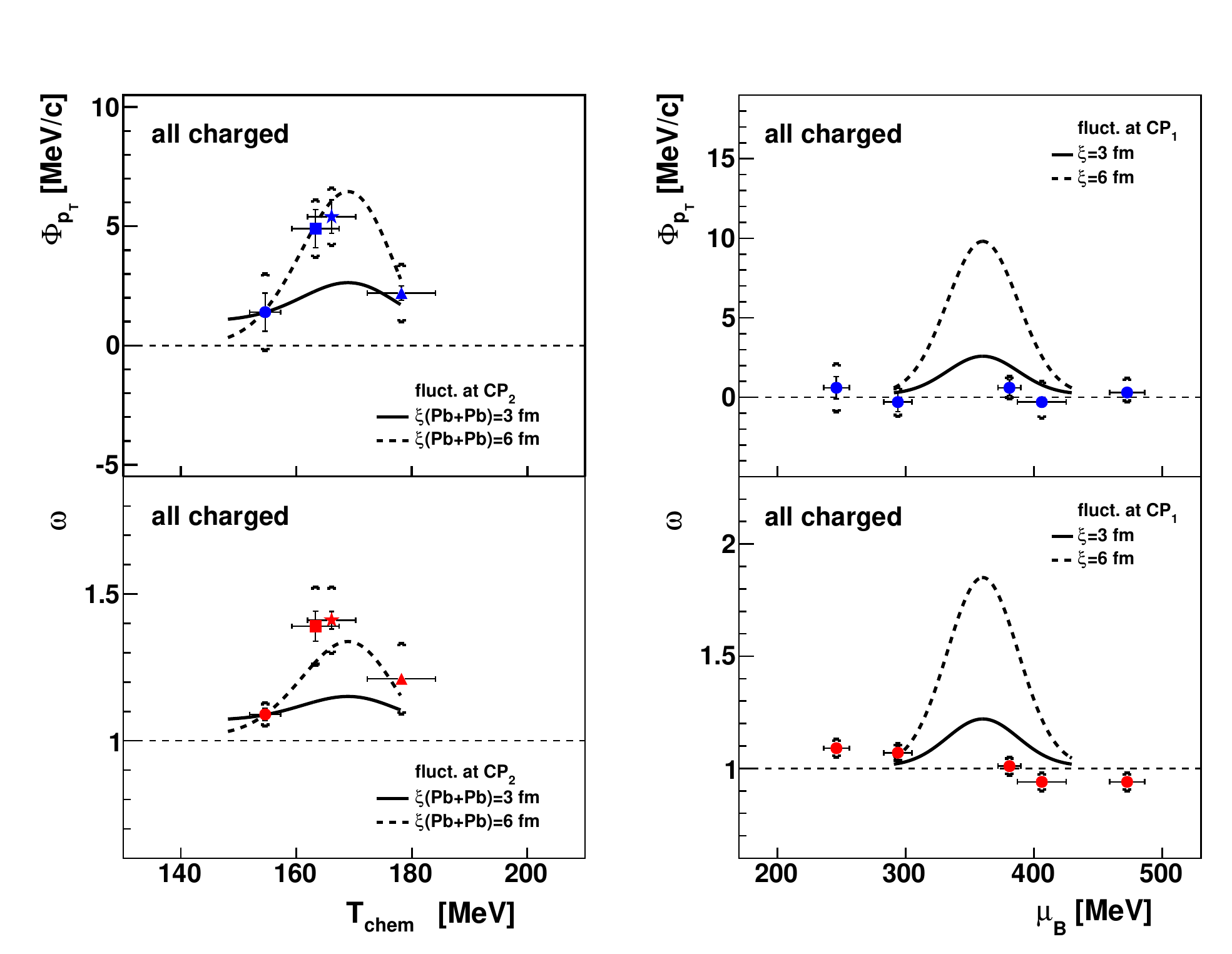}
\end{minipage}
\caption{\label{cp_signals}
System size dependence of the $\Phi$
measure of transverse momentum fluctuations ($top$) 
and scaled variance of multiplicity fluctuations ($bottom$)  for  charged
hadrons in central collisions of two identical nuclei.
$Left:$ Results for p+p interactions and central C+C, Si+Si and
Pb+Pb collisions at 158$A$~GeV are plotted as a function
of chemical freeze-out temperature~\cite{Be}.
$Right:$ Results for central Pb+Pb collisions at 20$A$, 30$A$, 40$A$,
80$A$ and 158$A$~GeV are plotted as a function of baryonic chemical
potential at chemical freeze-out~\cite{Be}.
The lines present  predictions for two
hypothetical locations of the critical point 
CP1 (T~=~147~MeV, $\mu_B$~=~360~MeV) (right plot)  and
CP2 (T~=~178~MeV, $\mu_B$~=~250~MeV) (left plot).
}
\end{center}
\end{figure}

The transition region can be studied experimentally in
nucleus-nucleus collisions only at $T$,~$\mu_B$ values which
correspond to collision energies higher than the energy of the
onset of deconfinement. This important conclusion is easy to
understand when looking at Fig.~\ref{cpod}. Signals of the critical
point can be observed provided the freeze-out point is close to it
(see Fig.~\ref{cpod}~{\it right}). On the other hand, by definition
the critical point is located on the transition line. Furthermore,
the energy density at the early stage of the collision is, of course,
higher than the energy density at freeze-out.
Thus, the condition that the freeze-out point is near the
critical point implies that the early stage of the system
is above (or on) the transition line. This in turn means that
the optimal energy range for the search for the critical point
lies above the energy of the onset of deconfinement
(see Fig.~\ref{cpod}~{\it left}).
This general condition limits the search for the critical point to
the collision energy range $E_{LAB} > 30A$~GeV.

The analysis of the existing experimental data~\cite{Be}
indicates that the location of the freeze-out point in the phase
diagram depends on the collision energy and the mass of the
colliding nuclei. This dependence is schematically indicated in
Fig.~\ref{cpod}~{\it right}. 
NA49 pilot data on collisions of
medium and light mass nuclei suggest that signals of the critical
point are visible in C+C and Si+Si collisions at 158$A$~GeV~\cite{Kasia,inter}.
Example results on the dependence of fluctuations on system size    
at 158$A$~GeV and on energy in central Pb+Pb collisions are shown in Fig.~\ref{cp_signals}. 

These results motivate  a systematic search for the
critical point. Similar to the study of the properties of
the onset of deconfinement, a two-dimensional scan in collision energy
and size of the colliding nuclei is required.
As presented in Fig.~\ref{plans}  this scan was already stared
by NA61/SHINE and the complete set of data should be registered by the end 
of 2015.
The basic components of the NA61 facility were inherited from NA49.
Several important upgrades, in particular, the new and faster TPC read-out,
the new Projectile Spectator Detector~\cite{kurepin} and
the installation of He beam pipes, allow to collect data
of high statistical and systematic accuracy.

\noindent
Acknowledgements: \\
The NA61 Collaboration: This work was supported by  
the Hungarian Scientific Research Fund (grants OTKA 68506 and 79840),
the Polish Ministry of Science and Higher Education (grants 
667/N-CERN/2010/0 and N N202 484339),
the Federal Agency of Education of the Ministry of Education and Science
of the Russian Federation (grant RNP 2.2.2.2.1547), the Russian Academy of
Science and
the Russian Foundation for Basic Research (grants 08-02-00018 and 09-02-00664),
the Ministry of Education, Culture, Sports, Science and Technology,
Japan, Grant-in-Aid for Scientific Research (grants 18071005, 19034011,
19740162, 20740160 and 20039012),
the German Research Foundation (grant GA 1480/2-1)
Swiss Nationalfonds Foundation (grant 200020-117913/1) 
and ETH Research Grant TH-01 07-3. \\
\noindent
The NA49 Collaboration: This work was supported by
the US Department of Energy Grant DE-FG03-97ER41020/A000,
the Bundesministerium fur Bildung und Forschung, Germany (06F~137),
the Virtual Institute VI-146 of Helmholtz Gemeinschaft, Germany,
the Polish Ministry of Science and Higher Education (1~P03B~006~30, 1~P03B~127~30, 
0297/B/H03/2007/33, N~N202~078735,  N~N202~078738, N~N202~204638),
the Hungarian Scientific Research Foundation (T032648, T032293, T043514),
the Hungarian National Science Foundation, OTKA, (F034707),
the Bulgarian National Science Fund (Ph-09/05),
the Croatian Ministry of Science, Education and Sport (Project 098-0982887-2878)
and
Stichting FOM, the Netherlands.

%%%%%%%%%%%%%%%%%%%%%%%%%%%%%%%%%%%%%%%%%%%%%%%%%%%%%%%%%%%%%%%%%%%%%%%%%%%%%%%%%%%%%%%%%%%

%----------------------------------------------------------------


\begin{thebibliography}{99}

\bibitem{qgp}
%\cite{Ivanenko:1965dg}
%\bibitem{Ivanenko:1965dg}
  D.~D.~Ivanenko and D.~F.~Kurdgelaidze,
  %``Hypothesis concerning quark stars,''
  Astrophysics {\bf 1}, 251 (1965)
  [Astrofiz.\  {\bf 1}, 479 (1965)], \\
  %%CITATION = ASTKB,1,479;%%
%\cite{Itoh:1970uw}
%\bibitem{Itoh:1970uw}
  N.~Itoh,
  %``Hydrostatic Equilibrium of Hypothetical Quark Stars,''
  Prog.\ Theor.\ Phys.\  {\bf 44}, 291 (1970), \\
  %%CITATION = PTPKA,44,291;%%
%\cite{Cabibbo:1975ig}
%\bibitem{Cabibbo:1975ig}
  N.~Cabibbo and G.~Parisi,
  %``Exponential Hadronic Spectrum And Quark Liberation,''
  Phys.\ Lett.\   {\bf B59}, 67 (1975), \\
  %%CITATION = PHLTA,B59,67;%%
%\cite{Collins:1974ky}
%\bibitem{Collins:1974ky}
  J.~C.~Collins and M.~J.~Perry,
  %``Superdense Matter: Neutrons Or Asymptotically Free Quarks?,''
  Phys.\ Rev.\ Lett.\  {\bf 34}, 1353 (1975), \\
  %%CITATION = PRLTA,34,1353;%%
%\bibitem{Shuryak:1980tp}
  E.~V.~Shuryak,
  %``Quantum Chromodynamics And The Theory Of Superdense Matter,''
  Phys.\ Rept.\  {\bf 61} (1980) 71.
  %%CITATION = PRPLC,61,71;%%

\bibitem{GaGo}
%\cite{Gazdzicki:1998vd}
%\bibitem{Gazdzicki:1998vd}
  M.~Gazdzicki and M.~I.~Gorenstein,
  %``On the early stage of nucleus nucleus collisions,''
  Acta Phys.\ Polon.\  {\bf B30}, 2705 (1999)
  [arXiv:hep-ph/9803462].
  %%CITATION = APPOA,B30,2705;%%

\bibitem{evidence}
%\cite{:2007fe}
%\bibitem{:2007fe}
  C.~Alt {\it et al.}  [NA49 Collaboration],
  %``Pion and kaon production in central Pb+Pb collisions at 20A and 30A GeV:
  %Evidence for the onset of deconfinement,''
  Phys.\ Rev.\   {\bf C77}, 024903 (2008)
  [arXiv:0710.0118 [nucl-ex]].  

\bibitem{review}
%\cite{Gazdzicki:2010iv}
%\bibitem{Gazdzicki:2010iv}
  M.~Gazdzicki, M.~Gorenstein, P.~Seyboth,
  %``Onset of deconfinement in nucleus-nucleus collisions: Review for pedestrians and experts,''
  Acta Phys.\ Polon.\  {\bf B42}, 307 (2011)
  [arXiv:1006.1765 [hep-ph]].

\bibitem{cp}
%\cite{Stephanov:1998dy}
%\bibitem{Stephanov:1998dy}
  M.~A.~Stephanov, K.~Rajagopal and E.~V.~Shuryak,
  %``Signatures of the tricritical point in {QCD},''
  Phys.\ Rev.\ Lett.\  {\bf 81}, 4816 (1998)
  [arXiv:hep-ph/9806219].
  %%CITATION = PRLTA,81,4816;%%

\bibitem{kumar}
  L.~Kumar et al. [STAR Collaboration], Proceedings of Quark Matter 2011,
  May 2011, Annecy, France.

\bibitem{schukraft}
  J.~Schukraft et al. [ALICE Collaboration], Proceedings of Quark Matter 2011,
  May 2011, Annecy, France.

\bibitem{tim}
  T.~Schuster et al. [NA49 Collaboration], Proceedings of Quark Matter 2011,
  May 2011, Annecy, France.

\bibitem{terry}
  T.~Tarnovsky et al. [STAR Collaboration], Proceedings of Quark Matter 2011,

  May 2011, Annecy, France.
\bibitem{proposal}
 N. Antoniou et al. [NA61 Collaboration], CERN-SPSC-2006-034 (2006).

\bibitem{antoni}
 A.~Aduszkiewicz and T~Palczewski et al. [NA61 Collaboration],
 poster 153 at Quark Matter 2011, May 2011, Annecy, France.


\bibitem{wnm}
%\cite{Bialas:1976ed}
%\bibitem{Bialas:1976ed}
  A.~Bialas, M.~Bleszynski, W.~Czyz,
  %``Multiplicity Distributions in Nucleus-Nucleus Collisions at High-Energies,''
  Nucl.\ Phys.\  {\bf B111}, 461 (1976).

\bibitem{ssr}
%\cite{Stephanov:1998dy}
%\bibitem{Stephanov:1998dy}
  M.~A.~Stephanov, K.~Rajagopal and E.~V.~Shuryak,
  %``Signatures of the tricritical point in {QCD},''
  Phys.\ Rev.\ Lett.\  {\bf 81}, 4816 (1998)
  [arXiv:hep-ph/9806219].
  %%CITATION = PRLTA,81,4816;%%).
  
\bibitem{Kasia}
%\cite{:2009vy}
%\bibitem{:2009vy}
  T.~Anticic {\it et al.} [ NA49 and NA61/SHINE Collaborations ],
  %``Search for the QCD critical point at SPS energies,''
  PoS {\bf EPS-HEP2009}, 030 (2009).
  [arXiv:0909.0485 [hep-ex]].

\bibitem{inter}
%\cite{Anticic:2009pe}
%\bibitem{Anticic:2009pe}
  T.~Anticic {\it et al.} [ NA49 Collaboration ],
  %``Search for the QCD critical point in nuclear collisions at the CERN SPS,''
  Phys.\ Rev.\  {\bf C81}, 064907 (2010).
  [arXiv:0912.4198 [nucl-ex]].

\bibitem{Be}
%\cite{Becattini:2005xt}
%\bibitem{Becattini:2005xt}
  F.~Becattini, J.~Manninen, M.~Gazdzicki,
  %``Energy and system size dependence of chemical freeze-out 
  %in relativistic nuclear collisions,''
  Phys.\ Rev.\  {\bf C73}, 044905 (2006).
  [hep-ph/0511092].

\bibitem{kurepin}
 A.~Kurepin et al.,
 poster 617 at Quark Matter 2011, May 2011, Annecy, France.



\end{thebibliography}
\end{document}